\documentclass[a4paper,fleqn,longmktitle]{cas-sc}
\usepackage[authoryear]{natbib}
\usepackage{amssymb}
\usepackage{siunitx}
\usepackage[usestackEOL]{stackengine}
\usepackage{adjustbox}
\usepackage{float}
\usepackage[title]{appendix}
\UseRawInputEncoding
\newcommand{\degree}{\ensuremath{^\circ}}
\let\printorcid\relax %remove ORCID(s) footnote

\title[mode = title]{The influence of Equation of State on impact dynamics between Pluto-like bodies}

\begin{document}

\shorttitle{The influence of Equation of State on impact dynamics between Pluto-like bodies}
\shortauthors{Shimoni et al.}

\author[First]{Yonatan Shimoni}
\ead{yonatansmn@gmail.com}

\author[First,Second]{Oded Aharonson}[orcid=0000-0001-9930-2495]
\cormark[1]

\ead{Oded.Aharonson@weizmann.ac.il}
\cortext[cor1]{Corresponding author}

\author[Third]{Raluca Rufu}[orcid=0000-0002-0810-4598]
\ead{raluca@boulder.swri.edu}
\cormark[1]

\address[First]{Dept. of Earth \& Planetary Sciences, Weizmann Institute of Science, Rehovot, 76100, Israel}
\address[Second]{Planetary Science Institute, Tucson, AZ,  85719, USA}
\address[Third]{Southwest Research Institute, Boulder, CO, 80302, USA}


%\linenumbers
\maketitle

\begin{abstract}
Impacts between planetary-sized bodies can explain the origin of satellites orbiting large ($R>500$~km) trans-Neptunian objects. Their water rich composition, along with the complex phase diagram of water, make it important to accurately model the wide range of thermodynamic conditions material experiences during an impact event and in the debris disk. Since differences in the thermodynamics may influence the system dynamics, we seek to evaluate how the choice of an equation of state (EOS) alters the system's evolution. Specifically, we compare two EOSs that are constructed by different approaches: either by a simplified analytic description (Tillotson), or by interpolation of tabulated data (Sesame). Approximately $50$ pairs of Smoothed Particle Hydrodynamics impact simulations were performed, with similar initial conditions but different EOSs, in the parameter space in which the Pluto-Charon binary is thought to form (slow impacts between Pluto-size, water rich bodies). Generally, we show that impact outcomes (e.g., circumplanetary debris disk) are consistent between EOSs. Some differences arise, importantly in the production of satellitesimals (large intact clumps) that form in the post-impact debris disk. When utilizing an analytic EOS, the emergence of satellitesimals is highly certain, while when using the tabulated EOS it is less common. This is because for the typical densities and energies experienced in these impacts, the analytic EOS predicts very low pressure values, leading to particles artificially aggregating by a tensile instability.
\end{abstract}

% %\includegraphics{figs/grabs.pdf}

% \begin{highlights}
% \item Research highlights item 1
% \item Research highlights item 2
% \item Research highlights item 3
% \end{highlights}

%                         1. INTRODUCTION
%************************************************************************
\section{Introduction}\label{intro}

Giant impacts are common in young Solar Systems \citep[e.g.,][]{quintana2016frequency} and play an important role in forming and shaping planetary systems and bodies (e.g., \citealt{citron2015formation}, \citealt{chau2018forming}). Impact outcomes are diverse, ranging from a hit-and-run scenario to complete merger of the colliding bodies. The subsequent dynamics in post-impact systems are complex, and may lead to  different evolutionary paths for the systems. A possible result of an impact, common in slow graze-and-merge type events, is the formation of a circumplanetary debris disk, and in some cases, the capture of the surviving impactor, which results in a system composed of a primary, secondary, and a debris disk.  Material in the debris disk may be accreted by the primary body (and secondary, if it exists), escape the system, coalesce into larger clumps to form satellitesimals or moons, or form planetary rings. One system thought to originate by such an impact event is the Pluto-Charon binary system \citep{canup2005giant,canup2011giant}, in which all six bodies (including the additional smaller satellites) lie approximately in the same plane and have nearly circular orbits. It is suggested \citep{canup2011giant,sekine2017charon} that Charon was captured almost intact following the impact into an eccentric orbit close to Pluto, and then migrated to its present-day location as its orbit become circular. How the other small satellites attained their current locations and their unique mean motion resonance chain is still an open question, yet an origin from a debris disk seems to be the most plausible scenario for the system configuration \citep{kenyon2019pluto} due to the satellites' low eccentricities and inclinations, and their consistent direction of rotation.

The Pluto-Charon system is the most famous one among the Trans-Neptunian Objects (TNOs) population. However, such multiple-body systems are not unique in the trans-Neptunian region, as there are dozens of known binaries \citep{noll2008binaries} and a few triplets. Interestingly, following the discovery a satellite orbiting Gonggong \citep{kiss2017discovery}, it has been established that all large TNOs ($R>500$ km) are  part of a multiple-body system, with secondary-to-primary mass ratios in the range of $10^{-4}-10^{-1}$. The formation of such systems can be explained by a variety of giant impacts common in the trans-Neptunian region, often resulting in the capture of small intact clumps \citep{arakawa2019early}.

Giant impacts are commonly simulated using Smoothed Particle Hydrodynamics (SPH) codes, and rely heavily on the assumed EOS. Water, which is abundant in TNOs, poses a challenge to model, as it is known to exist in more than 10 different solid phases \citep{petrenko1999physics}. The many phases of water makes the construction of a comprehensive EOS over the broad range of thermodynamic quantities and phase transitions a difficult task. With this motivation, we seek to evaluate how the choice of EOS affects the dynamics in SPH impact simulations between water-rich bodies. We note that there are other numerical effects that are not considered here, such as rotational instabilities in rigid bodies \citep{speith2006improvements}, density discontinuities \citep{hosono2013density}, and the effect of resolution on SPH calculations \citep[e.g.][]{kegerreis2019planetary}.

%************************************************************************
%                         1.1 EQUATION OF STATE
%************************************************************************

\subsection{Equation of State}
Two common approaches to construct an EOS are by an analytic approximation, or by a tabulated, semi-analytic approach. The analytic EOS that is widely used is Tillotson \citep{tillotson1962metallic}, which was initially developed to model hypervelocity impact of metals at the US Air Force Special Weapons Center. It combines data from explosive shock-wave experiments and statistical models of atom structure (Mie-Gr\"{u}neisen, Thomas-Fermi and Thomas-Fermi-Dirac models) to construct analytic equations. Later works (e.g. \citealt{melosh1989impact}) expanded the Tillotson EOS to also include geologic materials. In the Tillotson EOS thermodynamic phase space is divided into a compressed region ($\rho>\rho_{0}$) and an expanded region ($\rho<\rho_{0}$), where $\rho, \rho_{0}$ are the density and zero-pressure density, respectively. The expanded region is further divided into three subregions, based on the material's internal energy $u$: (1) expanded cold state ($u<u_{\rm iv}$), (2) expanded hot state ($u>u_{\rm cv}$), which converges to ideal gas at low densities, and (3) an intermediate state ($u_{\rm iv}<u<u_{\rm cv}$), known as the mixed phase state. Here, $u_{\rm iv}$, and $u_{\rm cv}$ are the energies at incipient vaporization and complete vaporization, respectively. In \cite{tillotson1962metallic} the mixed phase state was governed by the same equation as the expanded cold state. To avoid pressure discontinuities, in modern approaches the mixed phase state is often assigned an alternative equation, which is simply a linear interpolation of the equations of the expanded subregions \citep{holian1989hydrodynamic}. The interpolation is purely mathematical, and does not represent a physical phase transition. Tillotson is convenient to implement in hydrocodes, is computationally fast, but suffers from the lack of phase transitions, and the exclusion of entropy and temperature. Thus using Tillotson may be problematic in some cases \citep[e.g.][]{stewart2020shock}, which will be explored further here.

The alternative to analytic EOS approximations are tabulated data. Although computationally slower and of finite sampling in thermodynamic phase-space, the tabulated EOS is often more accurate. To cover the wide range of thermodynamic quantities, multiple models of atomic structures are implemented (e.g. some Sesame EOS \citep{lyon1992sesame} uses Saha ionization model to calculate the electronic contribution to the EOS; Einstein model to calculate lattice vibrations in solids). Other advantages to common tabulated EOSs are that they typically treat phase transitions and include all thermodynamic variables.

Though both approaches are widely used, there has been no attempt to thoroughly compare impact simulations between water-rich bodies, with similar geometric and dynamic initial conditions, but with different EOS. \cite{emsenhuber2018sph} studied the effects of different EOS (M-ANEOS vs. Tillotson) in Mars-forming impacts, i.e. of large ($R_{\rm{target}}\sim2000$ km) differentiated bodies with an iron core and a rocky mantle, with impact velocities of 4 km/s. They report small dissimilarities in the final body thermodynamic properties between the EOSs, but overall concluded that both EOSs produced similar results. Here, impacts and post-impacts debris disks in the early TNOs population are studied, on a wide impact parameter space, where melting may be substantial, in order to quantify the typical differences resulting between the EOSs. We present results from a set of $\sim50$ pairs of SPH impact simulations. 

%************************************************************************
%                          2. METHODS
%************************************************************************

\section{Methods}\label{s:methods} 

Planetary-scale impacts are usually simulated using the mesh-free Lagrangian method, SPH. SPH treats the material as self-gravitating fluid, meaning particles have no tensile strength and the governing equations are those of fluid dynamics (\textit{Navier-Stokes} for an inviscid fluid without thermal conductivity, but with artificial viscosity, to damp subsonic turbulence that could potentially propagate to large discontinuities). The density of a particle, $i$, is given by a 3D spline kernel ($W$), which is a function of distance and a defined smoothing length ($h$):

\begin{equation}
\rho_{i}=\sum_{j=1}^{N}m_{j}W(|\boldsymbol{r}_{ij}|,h_{i}),
\end{equation}
where $N$ is the number of particles in the simulation and $\boldsymbol{r}_{ij}\equiv \boldsymbol{r}_i-\boldsymbol{r}_j$ is the distance vector between particles $i$ and $j$. The particle€™s acceleration due to hydrodynamic forces is then given by \citep{springel2005cosmological}:

\begin{equation}\label{eq:1}
\frac{dv_{i}}{dt}=-\sum_{j=1}^{N}m_{j}\left[f_{i}\frac{P_{i}}{\rho_{i}^{2}}\nabla_{i}W_{ij}(h_{i})+f_{j}\frac{P_{j}}{\rho_{j}^{2}}\nabla_{i}W_{ij}(h_{j})+\Pi_{ij}\nabla_{i}\overline{W}_{ij}\right],
\end{equation}
where $m_j$ is the particle's $j$ mass, the coefficient $f=f(\rho,h)$ arises from the constraint that a fixed mass is contained within a smoothing volume, $W_{ij}(h)=W(|\boldsymbol{r}_{ij}|,h)$, $\Pi_{ij}$ is the artificial viscosity term, $\overline{W}_{ij}\equiv [{W}_{ij}(h_{i})+ {W}_{ij}(h_{j})]/2$, and the self-gravity term is not shown in equation (as customary). Angular and linear momenta are conserved as the force between pairs of particles is symmetric \citep{monaghan1992smoothed}.  The particles' dynamical and thermodynamical properties evolve through gravitational interactions, compressional heating, and shock dissipation.

\begin{figure}[pos=h]
\centering
  \vspace{-0.4cm}
  \includegraphics[scale=0.5]{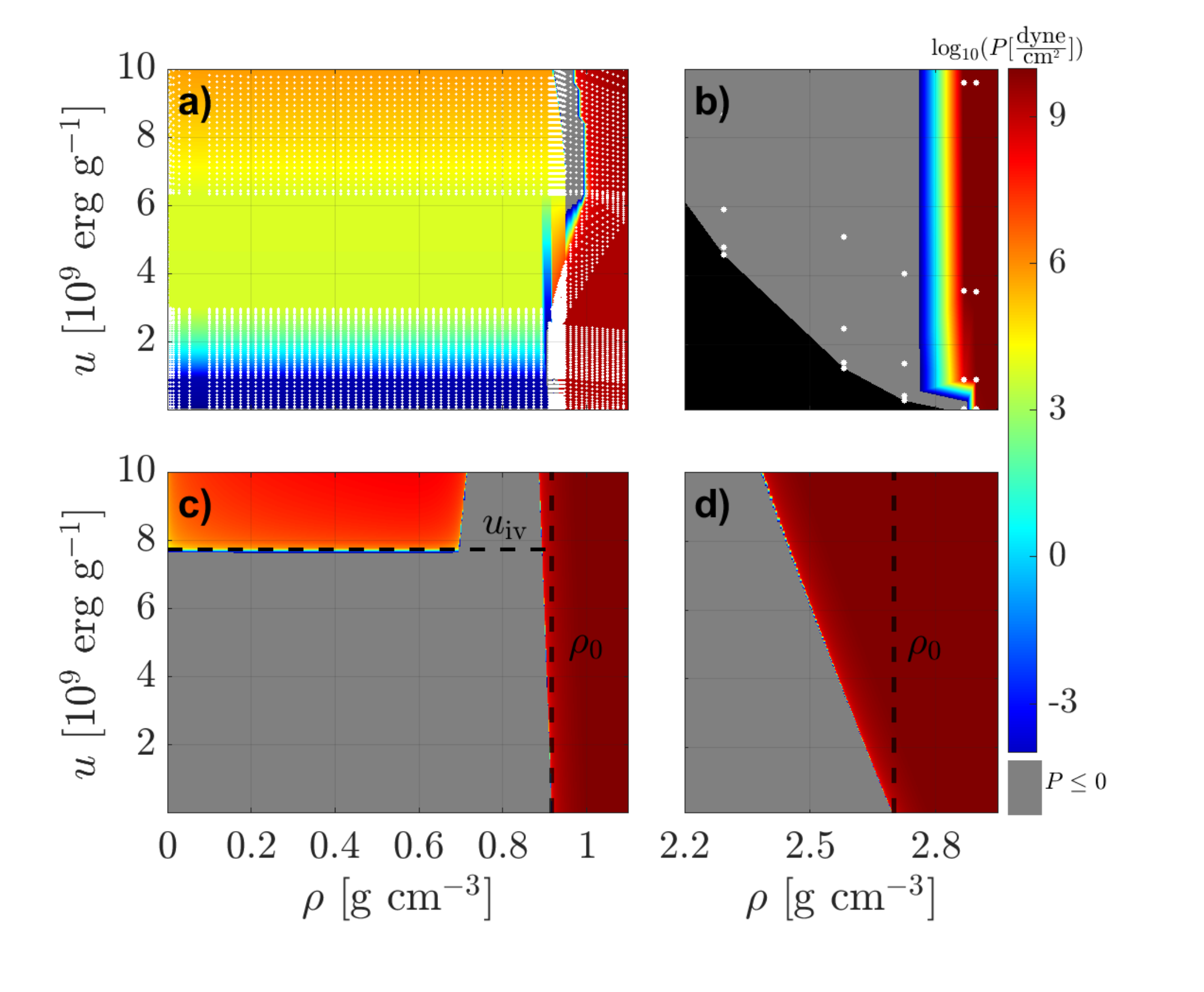}
  \vspace{-0.8cm}
  \caption{Phase-space diagrams of Sesame (a - water, b - basalt) and Tillotson (c - water, d - basalt) in the typical parameter space of Pluto-like bodies. The white dots in a) and b) represent the sampled points, whereas the pressure is interpolated. The black area in (b) indicates the absence of data and that thermodynamic properties are not defined in that region. The black dashed lines in c) and d) represent the Tillotson region's boundaries. Typically the rocky material particles in these simulations are in the compressed and expanded cold regions, whereas the water may also be in the mixed-phase region.}
  \label{f:eos}
\end{figure}

We have chosen Sesame to model basalt \citep{barnes1988sesame}, and the 5-phase EOS for water \citep{senft2008impact}. For the Tillotson parameters of basalt and water we used data from \cite{benz1999catastrophic}. Phase-space diagrams of the selected EOSs are shown in Fig.~\ref{f:eos}. At the low internal energy and density region of phase space, the Tillotson formulation results in negative pressures. Ejected particles can reach this region as their distances from other particles grow, and thus their densities reduce. In this case the particles' pressure is assigned a small value of $\sim 10^{-9}~ [\rm{dyne}~\rm{cm^{-2}}]$ for Sesame, and zero for Tillotson \citep{benz1999catastrophic}. We verified the simulation dynamics are not influenced by the choice of small value. A particle's ability to resist gravitational attraction from other particles depends on its pressure (Eq.~\ref{eq:1}), and so negative, or near-zero, pressure regions give rise to what is known as a "tensile instability" \citep{sigalotti2008adaptive,price2012smoothed}, which causes particles to artificially clump together. We therefore expect that when using Tillotson EOS in impacts between Pluto-like bodies, artificial clumps will form more readily in water-rich disks.

\begin{figure}[pos=h]
\centering
  \includegraphics[scale=0.5]{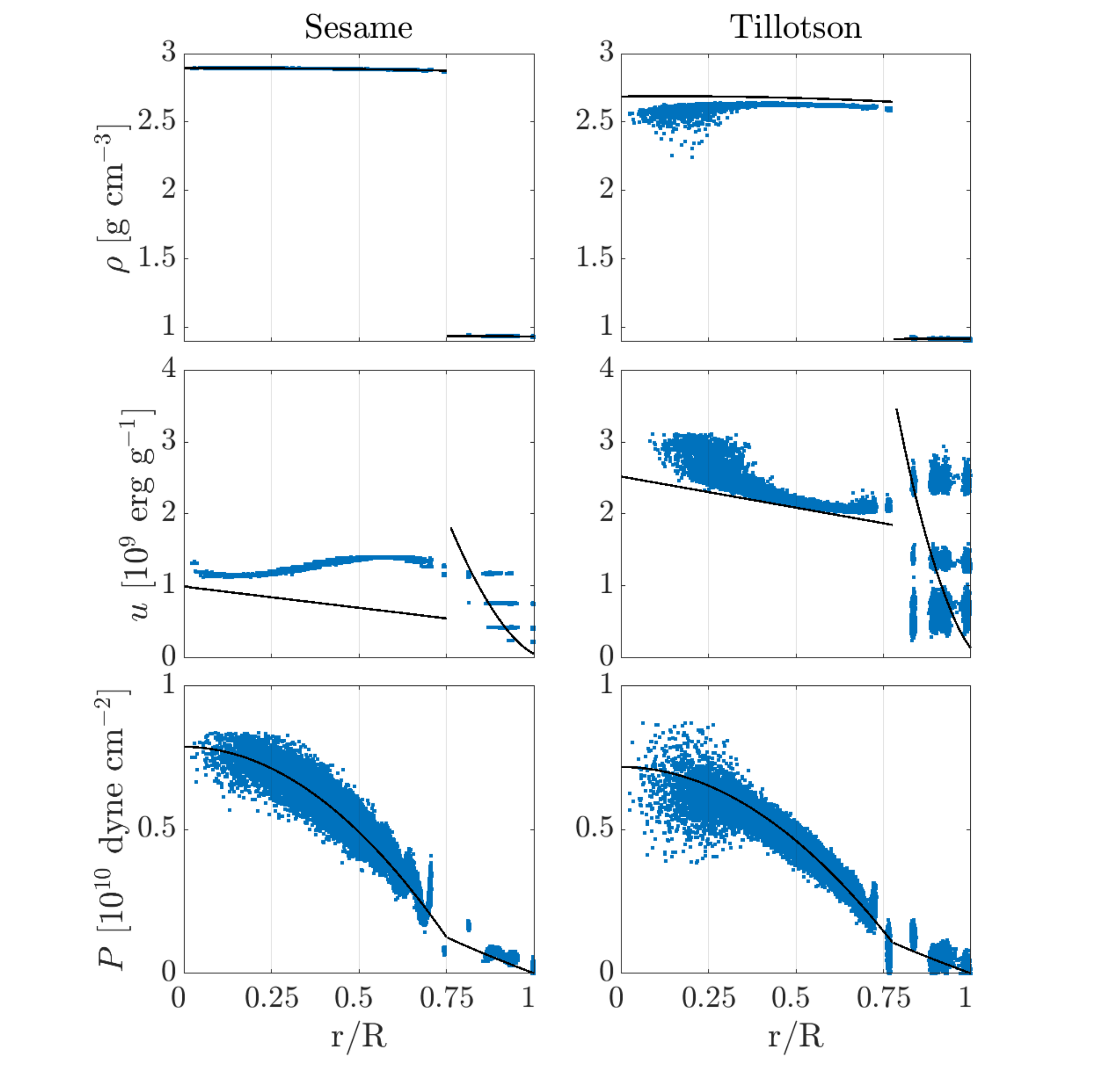}
  \caption{Initial profiles of a non-rotating body with 0.5 Pluto mass for Sesame (left) and Tillotson (right). Solid lines are the analytically calculated initial profiles, and the blue points are the particles' values after bodies were simulated in isolation and have reached hydrostatic equilibrium.}
  \label{f:initial_Profiles}
\end{figure}

%%% Bodies' Initialization %%%
The first step in an SPH simulation is to initialize the thermodynamical state of stable bodies. In an iterative process, the hydrostatic and mass conservation equations are solved, yielding the body's mass, radius and thermodynamic profiles. The initial temperature was set to $40$K at the surface, $240$K at the core-mantle boundary, and $300$K at the center, with a linear profile in between, and kept constant throughout iterations. To convert from initial temperature to  internal energy in Tillotson EOS, we use the specific heat capacity. The radii of the body and core-mantle boundary were set to obtain the desired body mass, and such that the average density matches Pluto's \citep{nimmo2017mean}. This resulted in core to mantle mass ratios dependent on the body's total mass, in the range of $M_{\rm core}/M_{\rm mantle}\approx2.5-3$. Particles were placed in a sphere using SEAgen algorithm \citep{kegerreis2019planetary} and then simulated with SPH in isolation until they reach thermodynamic equilibrium. The differences between the EOSs (Fig.~\ref{f:eos}) results in bodies with different initial energy and density profiles (Fig. \ref{f:initial_Profiles}). Water particles have similar densities in both cases, but basalt particles in the Tillotson formulation have a lower density.

%%% Parameter Space %%%
Impacts are described by two parameters: the angle ($\xi$) between the surface normal and the impact trajectory, and the impact velocity $v_{\rm imp}$, normalized to the mutual escape velocity, $v_{\rm esc} =\sqrt{2GM_{\rm T}/(R_{1}+R_{2})}$ (where $G$ is the gravitational constant, $M_{\rm T}$ is the total colliding mass, and $R_{1},~R_{2}$ are the radii of the impacting bodies). Due to differences in the thermodynamic profiles (Fig. \ref{f:initial_Profiles}), initial bodies between EOSs will have slightly different radii and masses. We verified these differences are small, and $v_{\rm esc}$ varies by  $<0.1\%$ between EOSs. Simulations were performed using the hydrocode SWIFT \citep{schaller2018swift}, with ${\sim10^{5}}$ particles. The impact parameter space was motivated by previous simulations \citep{canup2005giant,canup2011giant} for the formation of the Pluto-Charon system. The total impacting  mass is $M_{\rm T} \approx M_{\rm PC}$ (where $M_{\rm PC}$ is the total mass of the Plutonian system), and the impactor to total mass ratio, $\gamma$, is  0.5 or 0.3. The impact is assumed to occur in a plane that includes the centers of mass of both bodies and the impact angle is varied among 0\degree, 30\degree, 45\degree, and 60\degree. The impact velocity assumed is relatively small, $1-1.1\ v_{\rm esc}$. The target body was assumed to have no initial spin, while the impactor was either initially non-rotating or spinning with a period of $T=5$ or $10$ hours about the $z$ axis. 
Simulations were stopped after 96 hours, the typical time for the central body to relax to a stable spherical shape.

%%% Disk and Clumps Analysis %%%
At the end of a simulation particles were classified according to their orbital energy. Classification was done by calculating their semi-major axes, eccentricities, and the equivalent circular semi-major axis (the value after the orbit has undergone circularization) assuming angular momentum conservation. The equivalent circular semi-major axis is given by $a_{\rm eq}=h_z^2/(GM)$, where $h_{z}$ is the specific angular momentum normal to the equatorial plane of the central body, and $M$ is the post-impact central body mass. If $a_{\rm eq}$ is smaller than the radius of the central body, the particle is considered to be accreted to the central body (typically less than 0.5\% the total mass in a simulation). Otherwise, bound particles were classified as disk particles, and unbound ones as escaping. The expected mass of a single satellite accumulating from a debris disk, $M_{\rm satellite}$, was calculated from angular momentum conservation based on lunar accretion models of \cite{ida1997lunar}, similarly to \cite{canup2005giant}:

\begin{equation}\label{eq:expected_satellite_mass}
M_{\rm satellite}=\frac{1.9L_{\rm disk}}{\sqrt{GMa_{\rm Roche}}}-1.1M_{\rm disk} 
\end{equation}
where $M_{\rm disk}$ and $L_{\rm disk}$ are the disk mass and angular momentum, respectively, and $a_{\rm Roche}$ is the location of the Roche limit (which was calculated using the mean density of present-day Charon, \citealp{nimmo2017mean}, under the assumption of a fluid satellite). The equation above is for a satellite that accreting at $1.2\,a_{\rm Roche}$. Extended disks have high angular momentum and may yield estimates for the satellite mass that is larger than the initial disk mass according to Eq.~\ref{eq:expected_satellite_mass}. In this case, a satellite is expected to accrete beyond $1.3a_{\rm Roche}$, and we limit the mass of the satellite to be that of the disk.

Because the number and size distribution of intact material ejected by the impact is of interest for subsequent satellitesimal formation, we developed an algorithm to detect post-impact clumps. Two particles were considered in contact if their mutual distance was smaller than the minimum of their combined smoothing lengths and some fixed value $d_{\rm max} = 200$ km. Clumps, or satellitesimals, were defined as at least 100 particles in pairwise contact, equivalent to $\sim 10^{-3}M_{\rm Pluto}$. The orbital elements of these satellitesimals were computed and studied. We define $q$ as the mass ratio of a satellitesimal and its primary.

In order to calculate the radial density profile of the disk, the integral of the 3D density kernel $W(r,\theta,z)$ (in cylindrical coordinates) over $z$ and $\theta$ is needed. The integral in $z$ is performed analytically to obtain $W_{2D}(r,\theta)$. The integral in $\theta$ does not have a simple analytical solution; to obtain an analytic expression for it we used two approximations. First, far from from the origin, we may approximate $rd\theta\approx{}dy$, where the $x$-axis is chosen to pass through the center of the particle.  Second, by numerical calculation, that the value of integral over $y$ is very well approximated by the normalized value of the integrand at $y=0$:

\begin{equation}\label{eq:W1Da}
    \int W_{2D}(r,\theta)rd\theta\ \approx \int W_{2D}(x, y)dy \approx W_{2D}(x, y=0)
\end{equation}

\begin{figure}[pos=h]
\centering
  \includegraphics[scale=0.4]{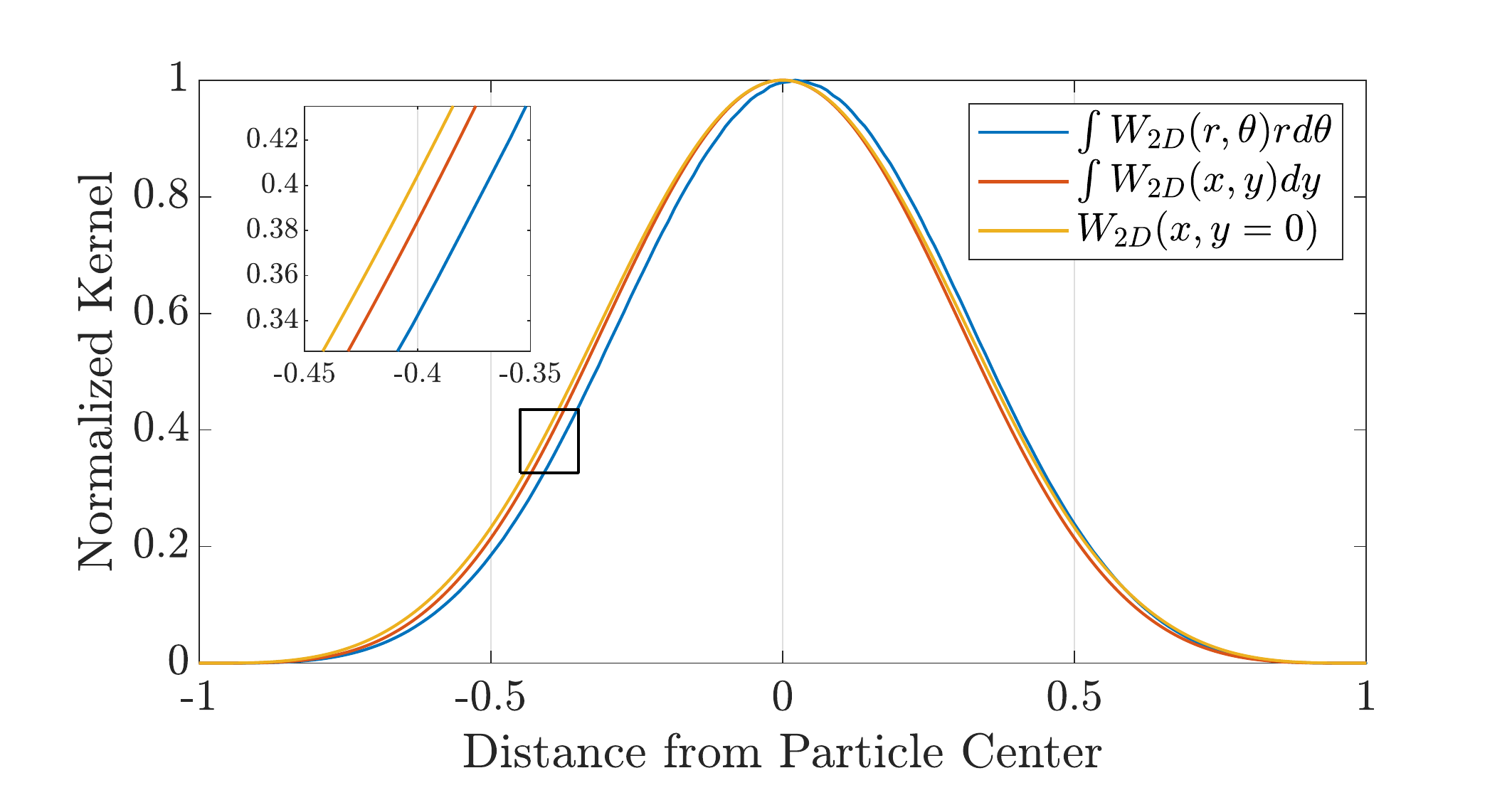}
  %\vspace{-1cm}
  \caption{2D kernel integrated over $\theta$, 2D kernel integrated over $y$, and a 2D kernel independent of $y$. Smoothing length is taken to be $1$.}
  \label{f:kernel}
\end{figure}

The validity of the approximations is seen in Fig.\ref{f:kernel}. The 1D kernel of a particle located at $r=2h$ is well approximated by the 1D kernel obtained by integration in $y$, and by the normalized 2D kernel evaluated at $y=0$.  Further from the origin the approximation improves.

%************************************************************************
%        3.1 RESULTS - NO SATELLITESIMAL FORMATION
%************************************************************************
\section{Results}\label{s:results}
In the following section we present analysis of cases with various parameters, summarizing the results of the simulations, and highlighting both consistency and inconsistencies between the two EOSs. Simulations are divided into several categories, according to their outcome. The bulk of the simulations is classified into three main categories, with a study case presented in each subsection: (\ref{results1}) In both EOSs a circumplanetary debris disk was formed, without the emergence of satellitesimals; (\ref{results2}) In both EOSs a circumplanetary debris disk was formed, with satellitesimals forming; (\ref{results3}) Inconsistency between EOSs in the formation of satellitesimals within a circumplanetary debris disk.

\subsection{Debris Disk with no Satellitesimal Formation} \label{results1}
\begin{figure}[pos=h]
\centering
\includegraphics[scale=0.6]{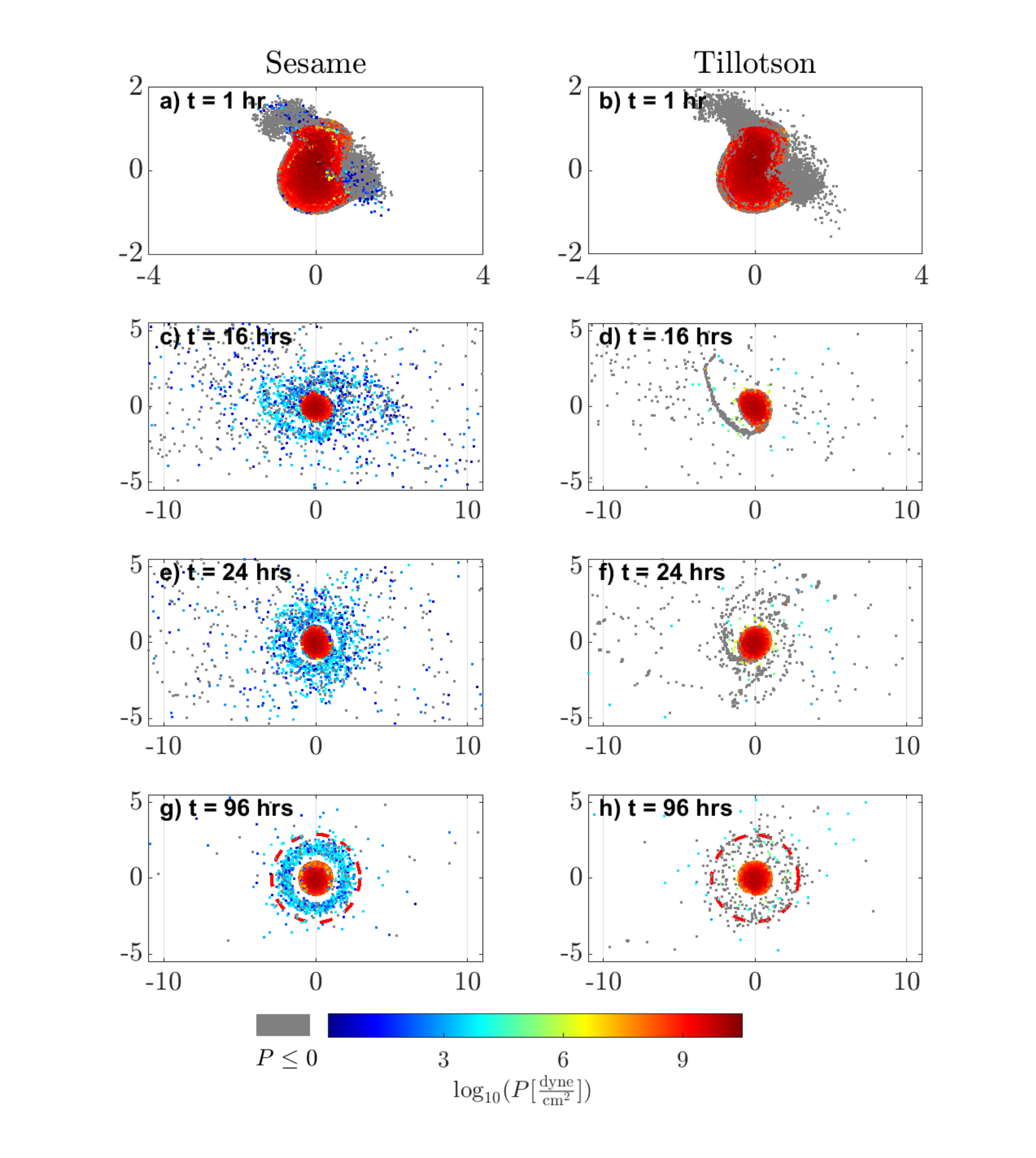}
\caption{Snapshots of a giant impact (left column - using Sesame EOS, right column - Tillotson EOS), between a rapidly rotating impactor ($T=5$ hours) and a relatively low impactor-to-total mass ratio ($\gamma=0.3$). This is an example of an accretionary impact with an impact angle of $\xi=30\degree$ in the plotted plane ($x,y$) and impact velocity  of $v_{\rm imp}= 1.1v_{\rm esc}$. The resulting debris disks have no satellitesimals at $96$ hours. Distances are normalized by Pluto's radius. Particles are projected on the equatorial plane, with one hemisphere removed for visibility of the impacting and final bodies (keeping particles with $z<0$).}
\label{f:DiskNoSat_FullSim}
\end{figure}

The results of a pair of simulations with impact angle $\xi=30\degree$, which did not produce satellitesimals are shown in Fig.~\ref{f:DiskNoSat_FullSim} in coordinates normalized by Pluto's radius $R_{\rm P}$. The system evolves similarly using the two EOSs, with more material being ejected from the bodies with Sesame, resulting in a more massive disk ($M_{\rm disk}/M_{\rm p}$=0.011 with Sesame, 0.009 with Tillotson. Subscript $p$ denote the primary). After a second collision, a long spiraling arm is created in both cases. Particle pressures differ along the arm between the EOSs: it is typically $\sim10^{4}[\rm{dyne~cm^{-2}}]$ with Sesame, while approaches zero in the Tillotson case. Satellitesimals do not form due to the low mass of debris, but a number of small particle concentrations appear, that do not pass our defined threshold (see Methods). Subsequently these structures break apart or escape the system, hence do not result in orbiting satellitesimals.

\begin{figure}[pos=h]
  \centering
  \includegraphics[scale=0.5]{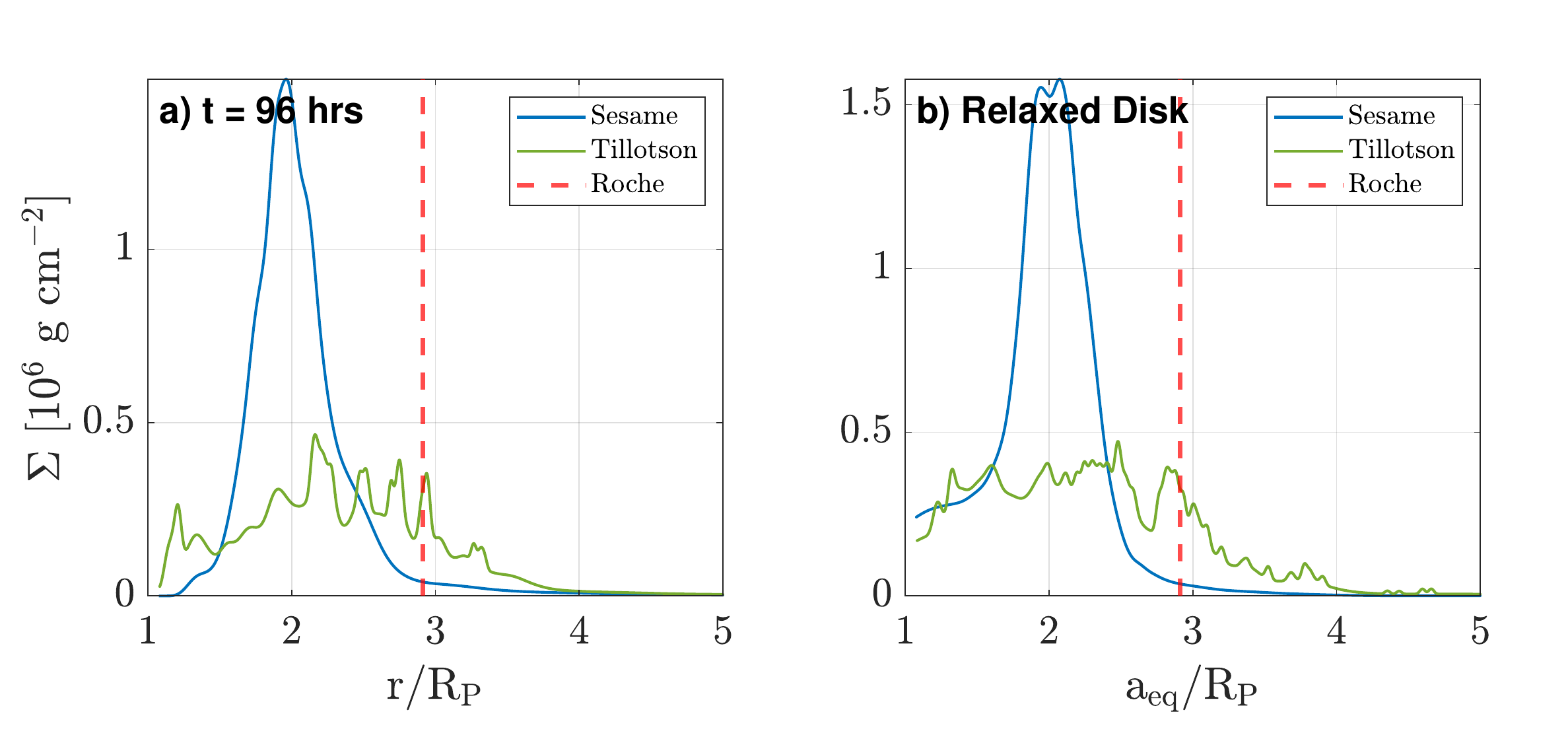}
  %\vspace{-1cm}
  \caption{a) Surface densities of the final disks in Fig.~\ref{f:DiskNoSat_FullSim} for Sesame (blue curve) and Tillotson (green curve) EOSs at the end of the SPH simulation (a) and after circularization (b). After circularization, both simulations result in a more confined disk, as expected.}
  \label{f:DiskNoSat_SurfaceDensity}
\end{figure}

%%% Disk dynamics and the state of the system at t_final %%%
The final debris disks differ in their radial extent and mass distribution. Most of the disk material in the Sesame run is confined within the Roche limit. Material in the Tillotson run is more dispersed, with several areas of local pressure enhancements. These differences are also evident in the averaged surface densities curves shown in Fig.~\ref{f:DiskNoSat_SurfaceDensity}a, with the Sesame EOS resulting in fewer peaks in the radial profile compared to the Tillotson run. In the Sesame case, the disk is concentrated near $r\sim2R_{\rm P}$, and drops off rapidly with only little mass outside the Roche limit. Surface densities were computed from the integrated kernels as described in the Methods section above. 

Following the impact, on a timescale of tens of orbits, the trajectories of particles are expected to circularize due to mutual collisions in the disk. This timescale is too long to study with SPH. Instead, to examine how these collisions modify the disk, the  semi-major axis equivalent of each particle's orbit was calculated by assuming that angular momentum is conserved as the eccentricity approaches zero. The resulting relaxed surface disk density (Fig.~\ref{f:DiskNoSat_SurfaceDensity}b) resembles that 
seen at the end of the SPH simulation (Fig.~\ref{f:DiskNoSat_SurfaceDensity}a). In both cases the disks maintain their approximate extent as well as their smooth (Sesame) and patchy (Tillotson) character. The circularization enhances the surface density in the inner part of the disk, as the equivalent semi-major axis is typically smaller than the particle's distance at the end of the SPH simulation.

Despite the large (factor of 2-3) difference in local surface densities, the final disks are similar in their total masses and specific angular momenta (Table \ref{table:DiskSatsProperties}), owing to the disk in Tillotson simulations having lower density in the inner portions, but extending to greater distances. Consequently, the expected secondary-to-primary mass ratios (calculated with Eq. \ref{eq:expected_satellite_mass}) is also similar. Disk compositions are similar as well, both being dominated by water ejected from the impactor and target's shallow mantles. As expected for slow,  moderate-angle impacts, the disks contain only a small fraction of the system's mass.

\begin{table}[pos=h]
    \begin{center}
    \begin{tabular}{ |c||c|c||c|c||c|c|| } 
    \hline
    & \multicolumn{2}{c||}{No Satellitesimals}     
    & \multicolumn{2}{c||}{With Satellitesimals}     
    & \multicolumn{2}{c||}{Satellitesimals in Till.} \\ \hline
    $\{\xi, v_{\rm imp}/v_{\rm esc}, \gamma, T~[\rm hrs]\}$
    &\multicolumn{2}{c||}{\{{30\degree, 1.1, 0.3, 5}\}}  & \multicolumn{2}{c||}{\{45\degree, 1, 0.3, $\infty$\}} &\multicolumn{2}{c||}{\{45\degree, 1.1, 0.5, $\infty$\}}\\ \hline
    EOS & Sesame & Tillotson & Sesame & Tillotson & Sesame & Tillotson \\
    \hline
    $M_{\rm {disk}}/M_{\rm p}$ & 0.011 & 0.009 & 0.029 & 0.037 & 0.124 & 0.089\\ 
    \hline
    $h_{\rm {disk}}/h_{\rm p}$ & 0.097 & 0.084 & 0.298 & 0.382 & 1.194 & 0.898 \\
    \hline
    $M_{\rm satellite}/M_{\rm p}$ & 0.006 & 0.006 & 0.029 & 0.037 & 0.106 & 0.089 \\
    \hline
    $N_{\rm{sats}}$ & - & - & 1 & 4 & - & 21  \\
    \hline
    $M_{\rm{sats}}/M_{\rm disk}$ & - & - & 0.481 & 0.460  & - & 0.436\\
    \hline
    Water Fraction & 0.89 & 0.99 & 0.36 & 0.43 & 0.99 & 0.96 \\
    \hline
    \end{tabular}
    \end{center}
    \caption{Disk and satellitesimal properties comparisons. The quantities reported are the normalized disk mass, normalized specific angular momentum $h$, secondary-primary mass ratio, number of formed satellitesimals ($N_{\rm sats}$), mass ratio of all satellitesimals combined ($M_{\rm sats}$) to the disk mass, and water fraction, using each EOS at the end of the simulations. In cases where satellitesimals formed, disk quantities include their properties. The table shows three cases: with no satellitesimals forming in either EOS, with satellitesimals forming in both EOSs, and with satellitesimals forming only in the Tillotson EOS. The parameters for each case are provided, infinite rotation period indicates no initial spin.}
    \label{table:DiskSatsProperties}
\end{table}

\begin{figure}[pos=h]
  \centering
  \includegraphics[scale=0.5]{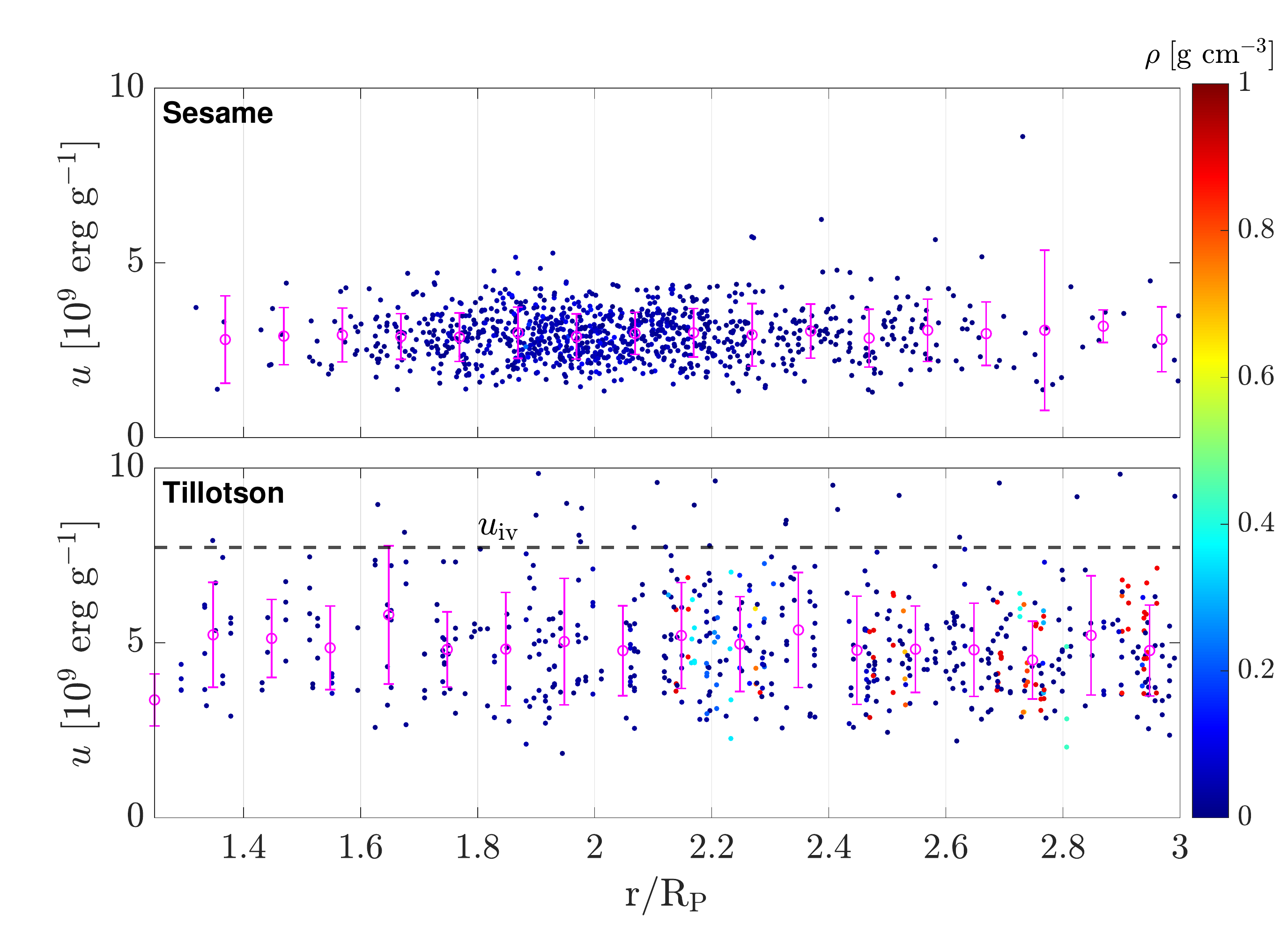}
   \caption{Internal energies of the final disks in Fig.~\ref{f:DiskNoSat_FullSim} using the Sesame (a) and Tillotson (b) EOSs. Mean and standard deviations in each radial bins is presented by the magenta marker and lines. Color corresponds to particles densities. The total disk mass beyond $3R_{\rm P}$ plotted here is negligible.}
  \label{f:DiskNoSat_u_prof}
\end{figure}

%%% Disk thermodynamics %%%
The water component, which dominates the debris disks, shows similar energy profile in its inner portions in both EOS cases (Fig.~\ref{f:DiskNoSat_u_prof}). The specific energy of individual particles can vary by a few $10^9~\rm{erg~g}^{-1}$, but when averaged, profiles show a roughly constant value with radius, indicating a roughly isothermal disk with temperature $\sim250$~K. According to Table~A1 in \cite{senft2008impact}, Sesame material, which exhibits extremely low densities, is in a mix state of ice Ih and vapor under these conditions. Material is mostly bound by $u=5 \cdot ~[\rm{erg~g^{-1}}]$, which at low densities corresponds to $T\sim273$K, the boundary above which ice Ih in the ice-vapor mixture transitions to liquid. Tillotson material shows larger variations in energy, and it is mostly below the energy of incipient vaporization. Given the low densities in the disk, Tillotson predicts the water particles to be either in a solid or liquid state but does not distinguish between the two \citep{BRUNDAGE2013461}.

%************************************************************************
%        3.2 RESULTS - SATELLITESIMAL FORMATION
%************************************************************************
\subsection{Debris Disk with Satellitesimal Formation} \label{results2}
\begin{figure}[pos=h]
\centering
  \includegraphics[scale=0.6]{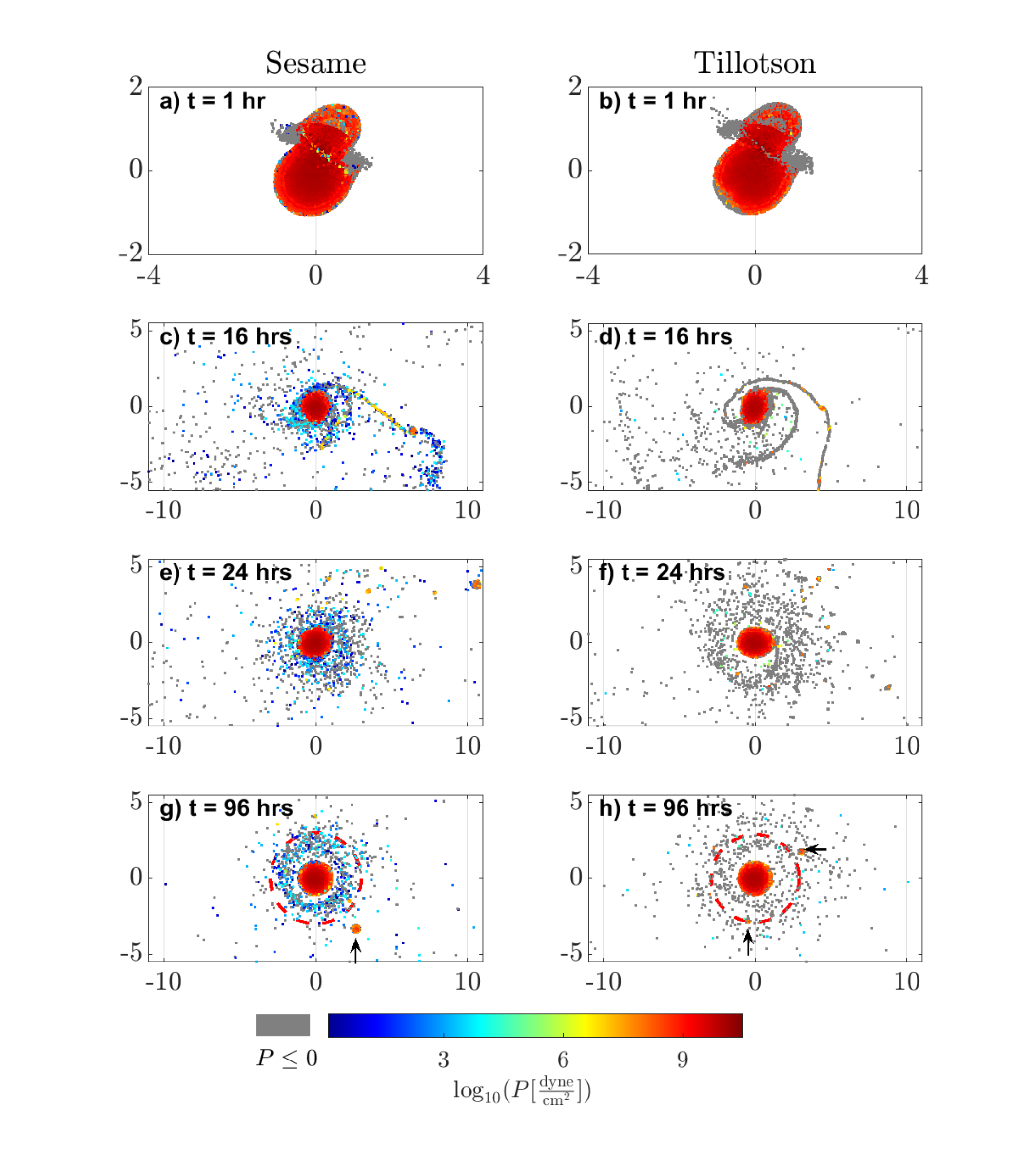}
  %\vspace{-0.6cm}
  \caption{Snapshots of an impact (left column - Sesame, right column - Tillotson) between a small non-rotating impactor and a relatively small impactor-to-total mass ratio ($\gamma=0.3$). The impact occurs at the escape velocity and at $\xi=45\degree$. Magenta arrows points to detected satellitesimals. Format is the same as Fig.~\ref{f:DiskNoSat_FullSim}.}
  \label{f:DiskWithSat_FullSim}
\end{figure}

%%% Brief intro %%%
A common outcome in our parameter space is the emergence of satellitesimals within the debris disk. Results of a pair of simulations are shown in Fig.~\ref{f:DiskWithSat_FullSim}. This is a graze-and-merge type impact, in which firstly the impactor grazes the icy mantle of the target, separates from the target and collides again within a few hours. In the secondary collision the bodies merge, and long spiraling ejecta arms are formed (\ref{f:DiskWithSat_FullSim}c, d), where satellitesimals and smaller clumps are created. In the Tillotson run, disk particles have unrealistically low pressures. Specifically, along the spiraling arms when a small clump is created it can easily attract neighboring particles. By the time the spiraling arms have almost completely broken apart (\ref{f:DiskWithSat_FullSim}f), many small clumps have already been formed, and by $t=96$ hours four satellitesimals coalesced, two of which are seen in \ref{f:DiskWithSat_FullSim}h.
In the Sesame run, pressures in the spiraling arms reach up to $10^{8}~[\rm{dyne~cm^{-2}}]$. One satellitesimal forms on an eccentric orbit (top right in \ref{f:DiskWithSat_FullSim}e), as well as some smaller clumps that will eventually fall inward and break apart in the disk.

\begin{figure}[pos=h]
  \centering
  \includegraphics[scale=0.4]{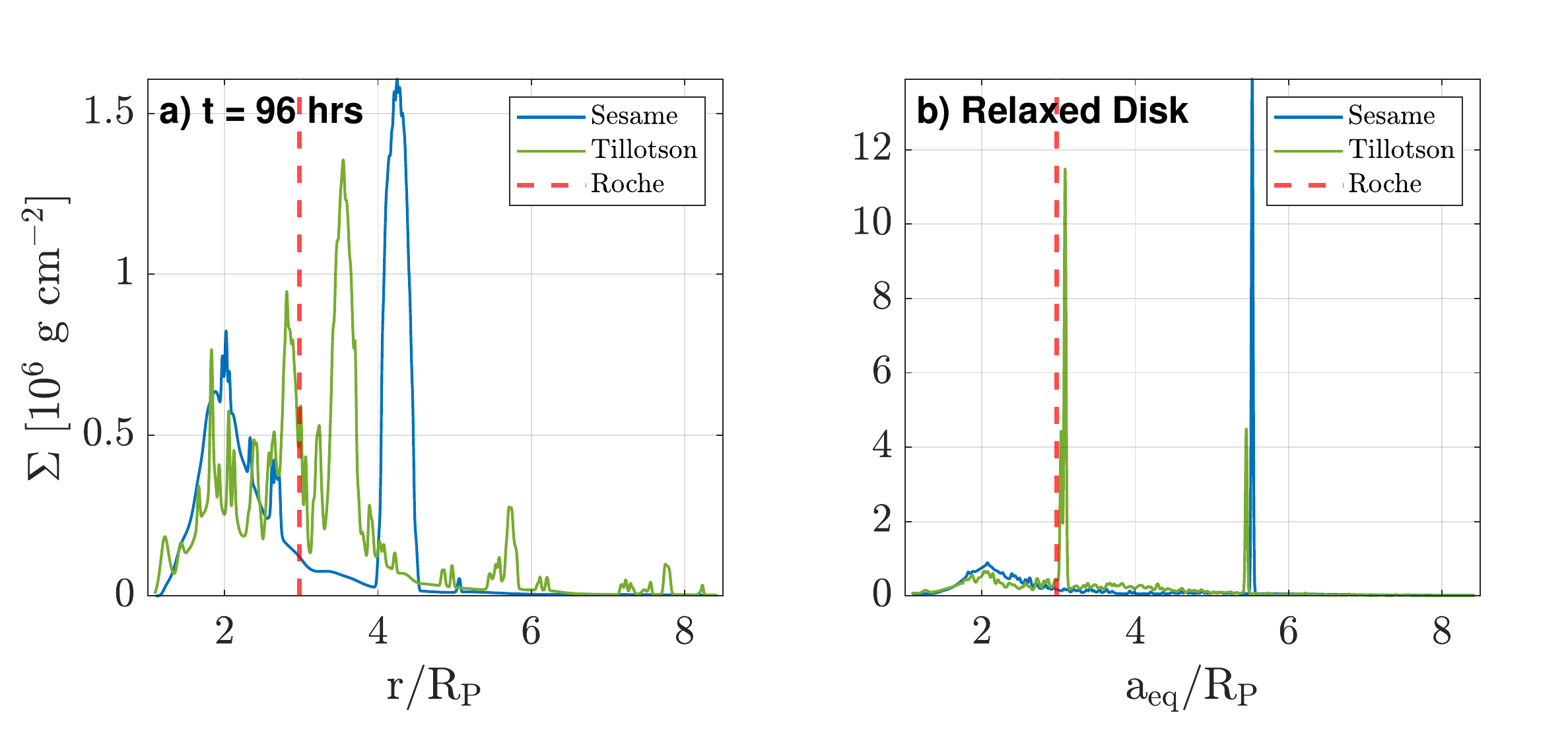}
  %\vspace{-1cm}
  \caption{a) Surface densities of the final disks in Fig.\ref{f:DiskWithSat_FullSim} for Sesame (blue curve) and Tillotson (green curve) EOSs at the end of the SPH simulation (a) and after circularization (b). After circularization, satellitesimals contain the majority of the mass in the disks, and the rest is mainly within the Roche limit.} 
  \label{f:DiskWithSat_SurfaceDensity}
\end{figure}

%%% Disk dynamics and the state of the system at t_final %%%
At the end of the simulations, disk material (excluding the satellitesimal) in the Sesame case is mostly located in an annulus interior to the Roche limit. The emerging satellitesimal's orbital elements are $e=0.67, a=10.07R_{\rm P}$ with mass ratio $q=0.014$. The debris disk in the Tillotson run extends to a greater distance. Two satellitesimals visible in the plot ($e=0.17$, $a=3.17R_{\rm P}$, $q=0.007$ and $e=0.11$, $a=3.07R_{\rm P}$, $q=0.005$), are located in proximity to the Roche limit, and two distant ones are at $37R_{\rm P}$ ($e=0.86$, $a=20.55R_{\rm P}$, $q=0.003$) and $83R_{\rm P}$ ($e>1$, $q=0.003$). The surface density profiles (Fig.~\ref{f:DiskWithSat_SurfaceDensity}), which were smooth in the case of no satellitesimals with Sesame, and undulating with Tillotson, have multiple distinct peaks with both EOSs. These localized concentrations of debris correspond to multiple satellitesimals.

At the end of simulations, the total satellitesimals mass is a significant fraction of the disks' mass (Table~\ref{table:DiskSatsProperties}). Following relaxation of the disks due to collisions (Fig.~\ref{f:DiskWithSat_SurfaceDensity}b) satellitesimals still account for most of the disk mass, with the remainder confined to within the Roche limit. At $t=96$ hours the two inner satellitesimals in the Tillotson case have intersecting orbits, and are expected to collide. Of the two distant satellitesimals in this simulation, one escapes the system, while the other settles in an orbit close to the central body (at $a_{\rm eq}=5.4R_{\rm P}$).

A summary of disk properties for the simulation described here is listed in Table~\ref{table:DiskSatsProperties}. The composition of the disks is rich in rocky material, mostly originating in the impactor's core. Rock fraction is comparable to Charon's \citep{stern2018pluto}, so a moon with a substantial rocky core could potentially accrete, but with mass smaller than that of Charon by a factor of $\sim5$.

\begin{figure}[pos=h]
  \centering
  \hspace*{-0.5cm}
  \includegraphics[scale=0.5]{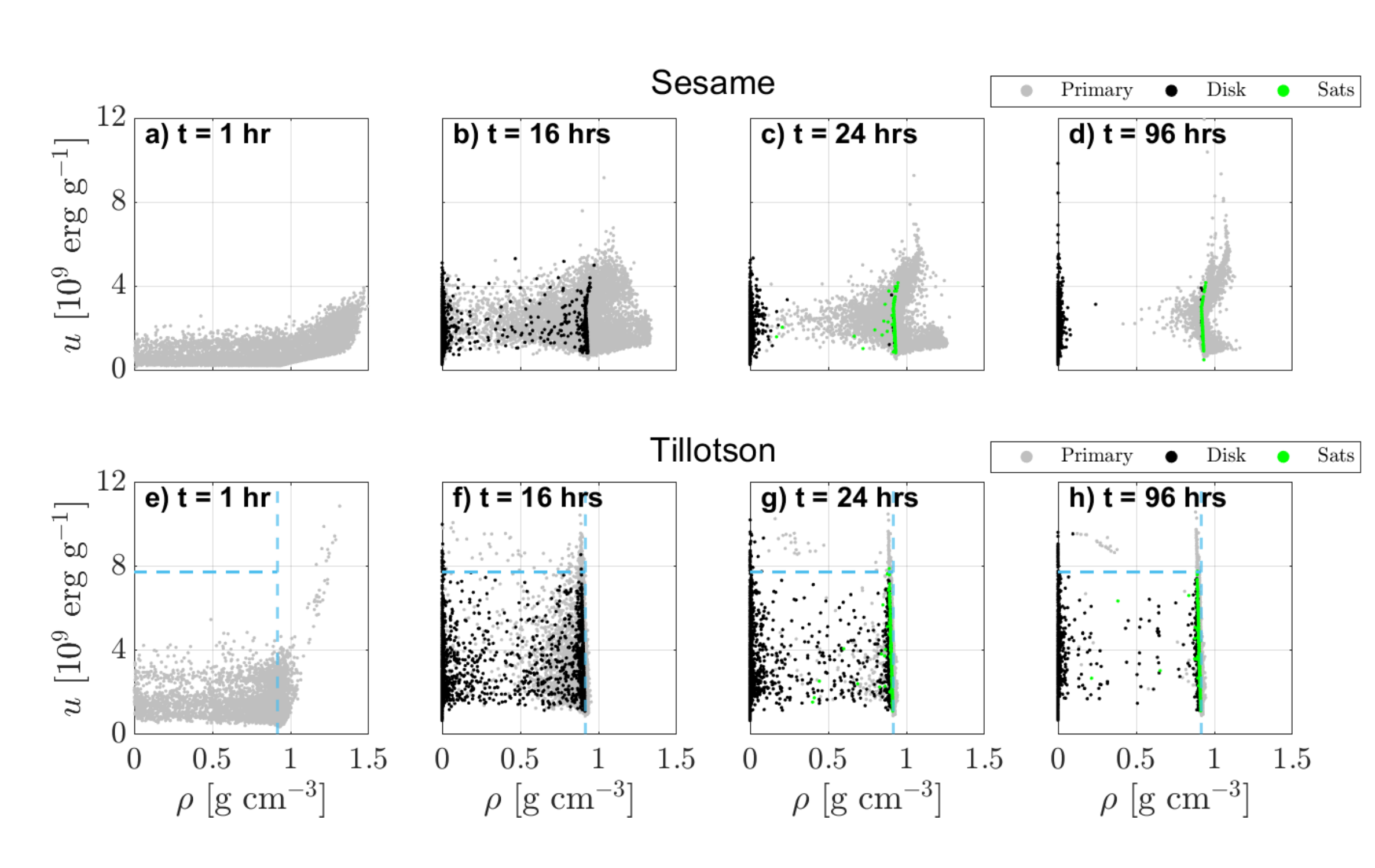}
  %\vspace{-0.75cm}
  \caption{Time evolution of the water particles' thermal state, for Sesame (top row) and Tillotson (bottom row). Boundaries of phases in Tillotson EOS are shown by dashed lines. The figure shows particles classified separately as primary and disk at $t\geq16$ hours, and satellitesimals are classified at $t\geq24$.}
  \label{f:DiskWithSat_thermal}
\end{figure}

%%% Disk thermodynamics%%%
Impact-induced thermodynamic conditions cover phases of water, from the familiar ice Ih, liquid, vapor to more exotic solid ice forms. It is therefore of interest to study how the two EOSs differ in their predictions. Fig.~\ref{f:DiskWithSat_thermal} shows the thermal state of water particles over time. Both runs occupy the same region in the thermodynamic phase space, yet water particles in the Sesame run reach higher densities (corresponding to high pressure ice phases), and are found in all classical phases and phase transitions in and after the impact. Water particles in the Tillotson run are either in the compressed region or the expanded cold region, and the majority of water particles in the Tillotson run have energies below the incipient vaporization energy. Satellitesimals water particles have densities similar to the zero-pressure density, {\it i.e.}, formed satellitesimal have an icy envelope. We note that in the SPH implementation, densities are influenced by the demand of having fixed mass within a smoothing length. Thus distant particles have large smoothing lengths, and hence reduced densities.

%************************************************************************
%        3.3 RESULTS - Satellitesimal only in Tillotson
%************************************************************************
\subsection{Disparities in the Number of Satellitesimals} 
\label{results3}
\begin{figure}[pos=h]
  \centering
  \includegraphics[scale=0.4]{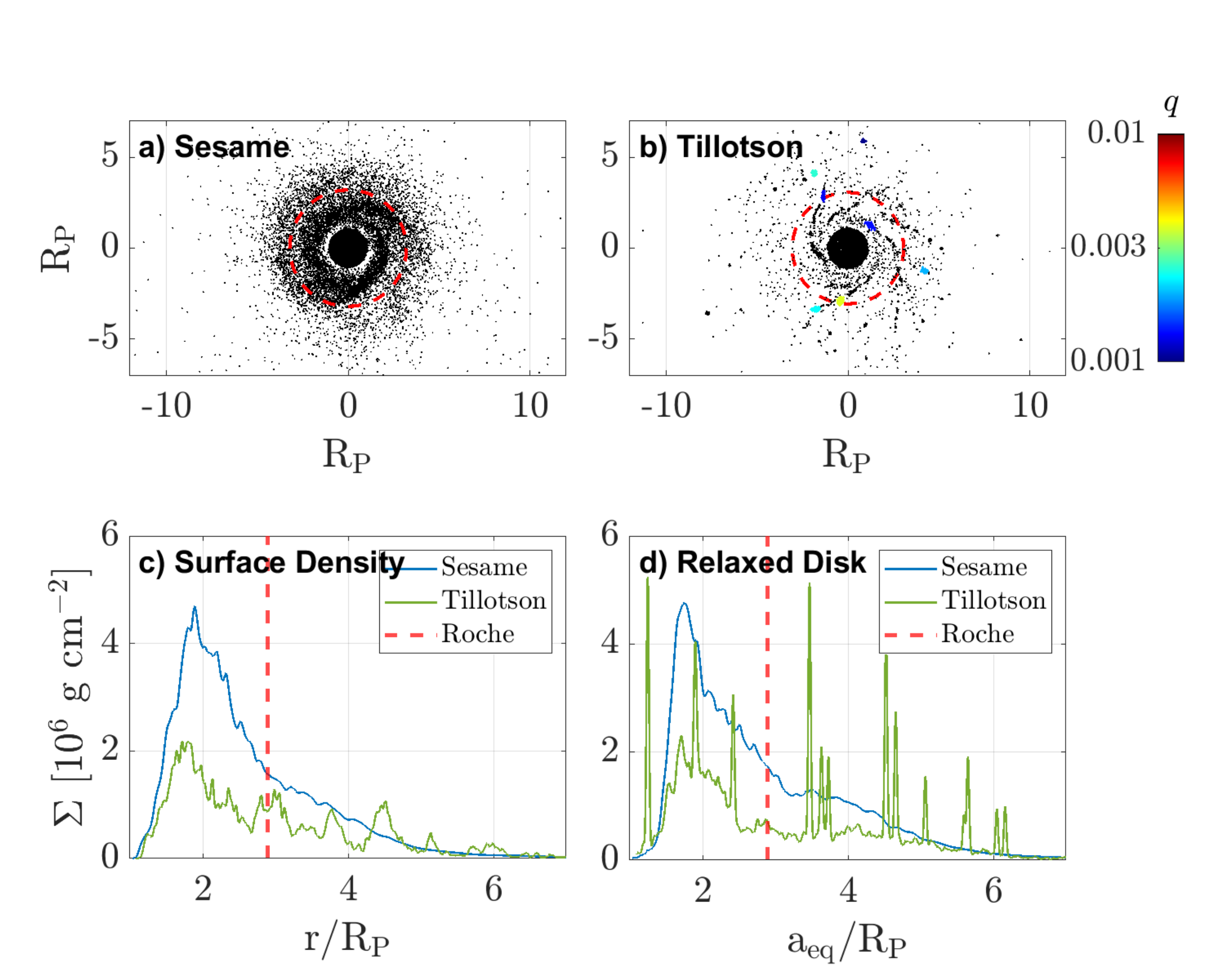}
  %\vspace{-0.5cm}
  \caption{The final state of systems with distinct results - no satellitesimal formation with Sesame and 21 satellitesimals with Tillotson (some beyond the plot limits). Format is the same as Fig.~\ref{f:DiskNoSat_FullSim}.}
  \label{f:DiskManySat}
\end{figure}

The previous section highlighted the problematic treatment of low pressure in the Tillotson disk, which in turn may lead to the aforementioned tensile instability. In the 12 pairs of simulations in which satellitesimals formed with one EOS but not the other, 11 were simulations using Tillotson. The only exception is when the impactor in the Tillotson run was almost entirely merged with the target body upon a secondary impact, and the resulting disk when using Sesame was a factor of $\sim3$ more massive than with Tillotson. In the following example (Fig.~\ref{f:DiskManySat}), an impact between non-rotating equal mass bodies at $45\degree$ and $v=1.1v_{\rm imp}$ is presented. This impact generated a relatively massive disk within our parameter space. The simulation with Sesame EOS yielded no satellitesimals, while the Tillotson run produced 21 of them (with a total mass of $0.04M_{\rm p}$), of which 15 are bound to the central body. The high number of satellitesimals is due to the large disk mass and the fact that it is composed almost completely of water particles, which are more easily clumped together artificially (due to the larger phase space in the Tillotson EOS that results in zero pressures). 

In the Tillotson simulation, a substantial amount of the disk material is coagulated in the satellitesimals and the rest of the disk is sparse. Seven of the bound satellitesimals have highly eccentric orbits ($e>0.77$), but upon orbit circularization they relax into a more compact disk ($9R$) where mutual collisions are expected. The resulting disk in the Sesame simulation does not contain any satellitesimals and in contrast to the Tillotson disk, a significant amount of its mass is inside the Roche limit ($55\%$). More mass within the Roche limit in simulations with Sesame is a general outcome in our simulations. As shown in Table \ref{table:DiskSatsProperties}, the resulting disk mass in  Sesame run is larger by a factor of $1.4$ compared to the Tillotson run but their overall composition is similar. 

%************************************************************************
%                       4. DISCUSSION     
%************************************************************************
\section{Discussion}\label{s:discussion}
Our results show how the choice of an EOS alters the predictions of giant impacts simulations in the regime of slow impacts between water-rich Pluto-size bodies. Two different approaches for constructing an EOS are compared - Tillotson (analytical) and Sesame (tabulated). Below we discuss and summarize our findings.

%************************************************************************
%        4.1 RESULTS - PARAMETER SPACE EXPLORATION
%************************************************************************
\subsection{Parameter Space Exploration}
\begin{figure}[pos=h]
  \centering
  %\hspace*{-1.75cm}
  \includegraphics[scale=0.4]{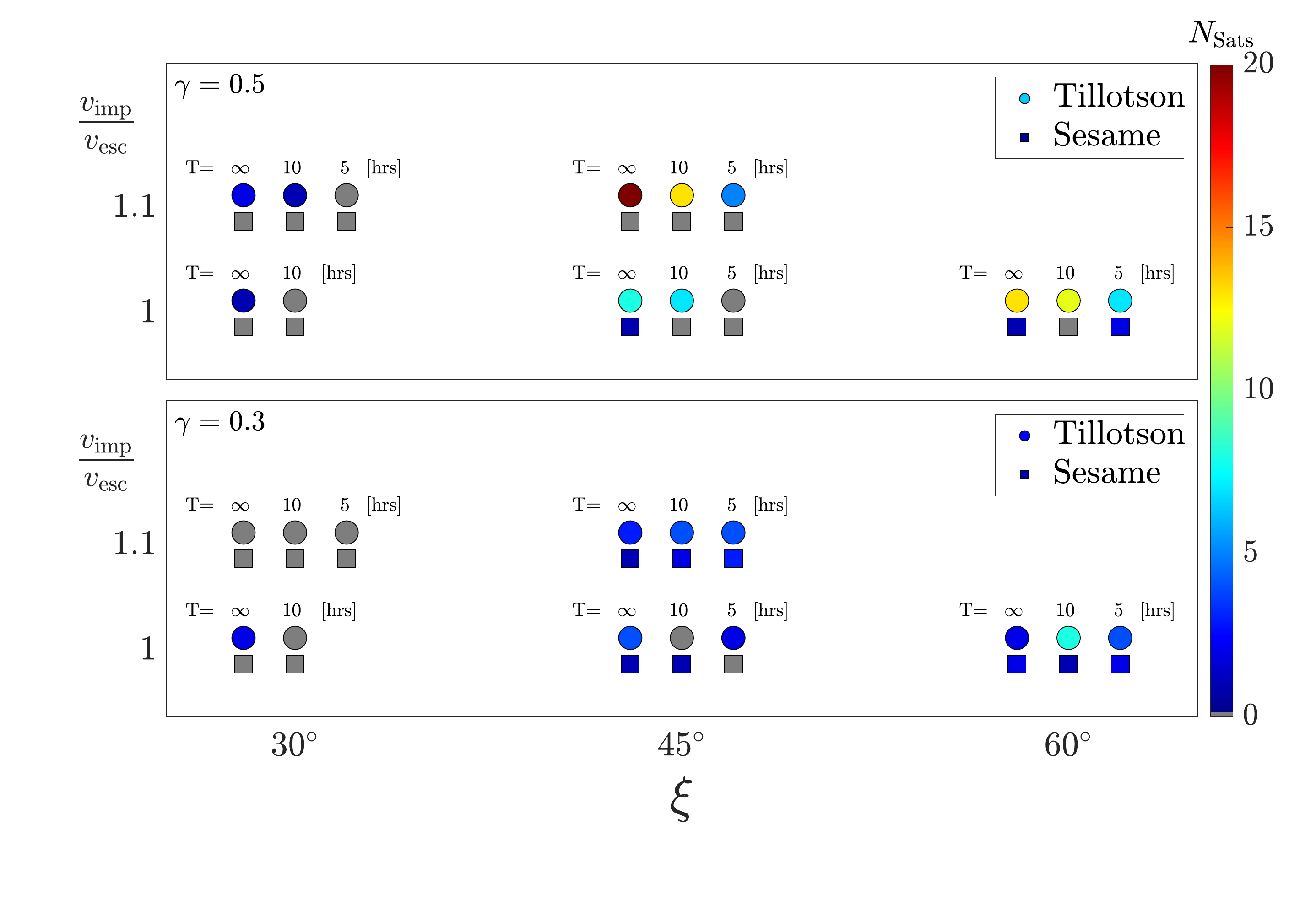}
  %\vspace{-1cm}
  \caption{Parameter space of disks-generating impacts. a) Impacts between equal mass bodies, and b) Impacts between a small impactor and a large target. Overall 140 satellitesimals were formed, 39 of which have periapses inside the Roche limit, and 45 are unbound to the central body. Head-on impacts did not formed debris disks and hence are excluded here.} 
  \label{f:ParameterSpace} 
\end{figure}

Simulations in our parameter space yield circumplanetary debris disks, and in most cases predict the formation of satellitesimals (Fig.~\ref{f:ParameterSpace}). In impacts between equal mass bodies, the interacting mass (the mass within the geometric cross section of the impact and target bodies, \citealt{leinhardt2011collisions}) is higher and more massive disks are generated than in an unequal mass ratio collisions. When using Sesame, dense disks are formed with most of the mass within the Roche limit. Satellitesimals form more commonly in lower mass ratio cases, such as in impacts with $\gamma=0.3$ because in these impacts more material is ejected beyond the Roche limit, and the inner disks in these cases are lower mass. Simulations with Tillotson form systems with multiple satellitesimals (sometimes more than 10). This is attributed to aggregation of material due to a tensile instability that occurs at low pressure \citep{price2012smoothed}. Further N-body simulations on longer timescales are required in order to test how such systems with multiple satellitesimals evolve. 

\begin{figure}[pos=h]
  \centering
 % \hspace*{-1.75cm}
  \includegraphics[scale=0.5]{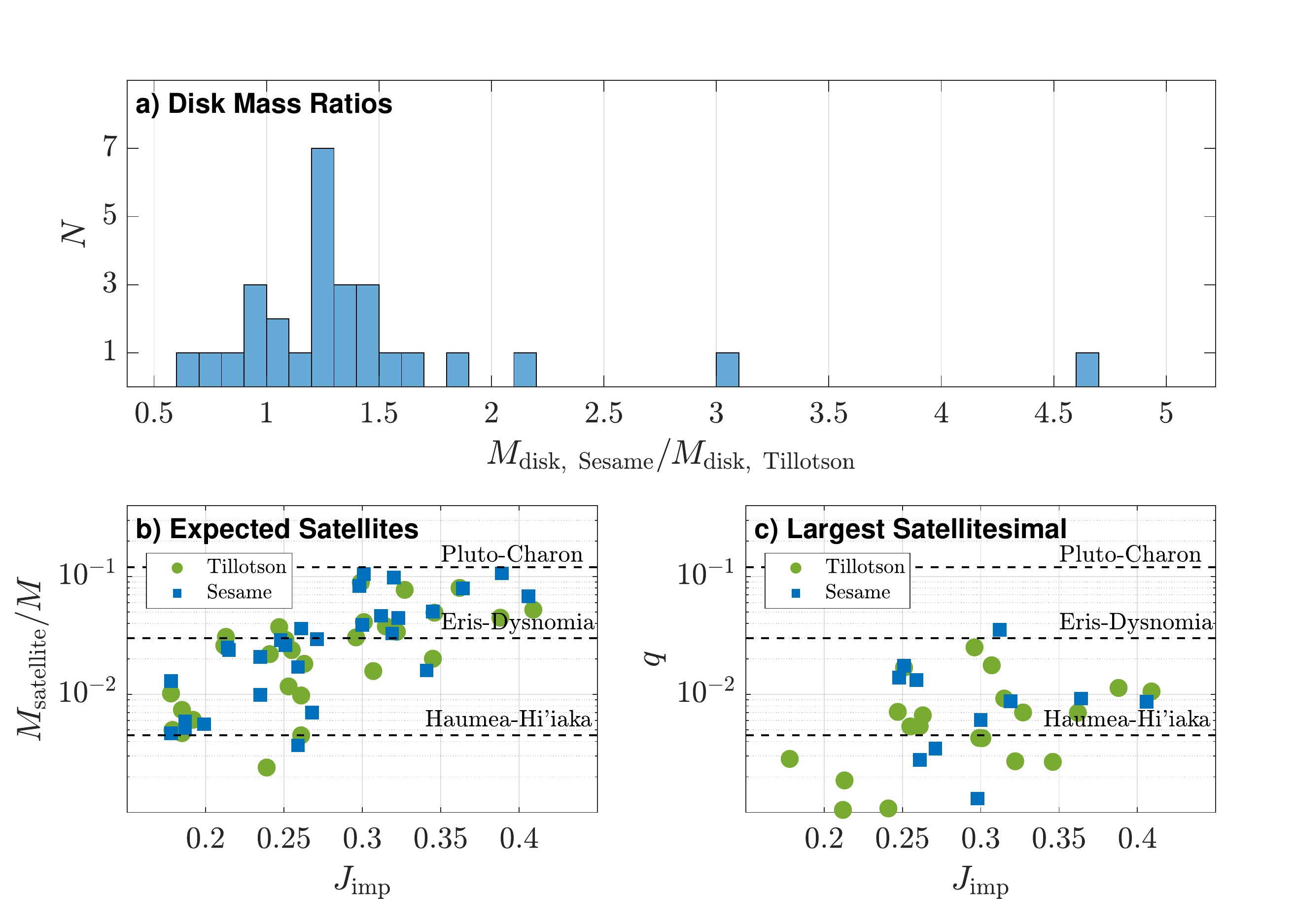}
  %\vspace{-1cm}
  \caption{a) Mass ratio of disks with identical initial conditions and different EOSs, b) The expected mass of a satellite accumulating from a disk, c) The mass of the largest formed satellitesimal in a simulation, plotted as a function of the normalized angular momentum at the moment of impact. Mass ratios of the three largest TNOs and their largest satellite are shown \citep{arakawa2019early}.} 
  \label{f:M_vs_Jimp} 
\end{figure}

Simulations using Sesame EOS generally predict disks which are more massive (Fig.~\ref{f:M_vs_Jimp}a) and have higher angular momentum than those using Tillotson. As a result, the mass of a single accumulating satellite is usually larger when using Sesame (Fig.~\ref{f:M_vs_Jimp}b) regardless of the impact parameters. Nevertheless, more massive disks do not correspond to larger satellitesimals (Fig.~\ref{f:M_vs_Jimp}c) nor to a greater number (Fig.~\ref{f:ParameterSpace}). The dependence of the size of the largest satellitesimal on the normalized angular momentum at the moment of impact $J_{\rm imp}$ (calculated as in \citealt{canup2005giant}) is weak, yet this parameter is important for the overall disk mass \citep{ida1997lunar}. The largest satellitesimal is relatively low in mass ({$q\sim 10^{-3}-10^{-2}$}), and is typically much less massive than a satellite expected to accumulate from the disk. However, the resulting disks lie mostly within the Roche limit, so their mass will partially accrete onto the primary, reducing the resulting satellite mass.

In terms of final configuration, both EOS approaches yield the same qualitative outcome for the same initial conditions. This is true over the range of our simulations. The resulting circumplanetary disks are similar in composition, and occupy the same region in the thermodynamic phase space, when comparing the two EOSs. 

Quantitatively, some differences emerge between EOSs, most notably the satellitesimals frequency distribution. Due to the tensile instability discussed in section~\ref{s:methods}, the use of Tillotson EOS enhances their formation, sometimes in numbers that exceed those of present-day systems. Tillotson handles water poorly and should be avoided in impacts between water-rich bodies. 
Correspondingly, satellitesimals forming in simulations using Tillotson are suspect, because disk pressures are artificially low (see Fig.~\ref{f:DiskNoSat_FullSim} and \ref{f:DiskWithSat_FullSim}).
More dissimilarities are seen in the disks' structures and radial extent. While disks in runs with Sesame have a dense annulus structure, disks in runs with Tillotson extend further (with greater distance separating particles). Satellitesimals properties are not consistent between runs with the same initial conditions but different EOSs. 

Two end-member cases in our parameter space are head-on impacts ($\xi=0\degree$ for all tested velocities), and oblique, faster than escape velocity impacts ($v_{\rm imp}/v_{\rm esc}=1.1$, and $\xi=60\degree$). For head-on impacts, a single merged body is formed, with the vast majority of ejected particles accreted by the planet, and the rest ejected to space. The final bodies are similar between EOSs. Impacts with $v_{\rm imp}/v_{\rm esc}=1.1$ and $\xi=60\degree$ resulted in two loosely bound (${e\geq0.98}$) or unbound bodies, with little mass transferred between the two or ejected to space. This result is consistent with previous studies of planetary impacts \citep[e.g.,][]{stewart2012collisions}, however in studies of impacts in the early TNO population \citep[e.g.,][]{arakawa2019early,canup2005giant,canup2011giant} for $\xi \geq 60\degree$, simulations resulted in more massive circumplanetary disks and intact moons found here. This difference may arise from the assumptions on the internal structure of the bodies, but further work is required to confidently identify the cause.

\subsection{Forming Moons of TNOs}
Our simulations offer the opportunity to examine formation of systems with multiple-bodies in the TNO population. Satellitesimals with masses comparable to those of known large TNO's satellites were successfully formed, whether as intact fragments following the impact or as expected accumulations from disk debris. In the former case, the largest mass we found is $3\%$ that of the primary (Fig.~\ref{f:M_vs_Jimp}c), while in the latter case, masses of the expected accreted satellites reaches $\sim$10\%, comparable to all observed satellites in the trans-Neptuninan region (Fig.~\ref{f:M_vs_Jimp}b). The exact number of intact satellites that form sensitively depends on initial conditions, and can be predicted approximately in across parameter space (Fig.~\ref{f:M_vs_Jimp}c).

None of the simulations produced an intact fragment with a mass comparable to that of Charon  (Fig.~\ref{f:M_vs_Jimp}c). Therefore, we conclude that within the examined parameter space, the Pluto-Charon pair did not form by an impact between fully differentiated bodies. An impact between partially differentiated bodies may be required for this system \citep{canup2005giant,canup2011giant}. 

\begin{figure}[pos=h]
  \centering
 % \hspace*{-1.75cm}
  \includegraphics[scale=0.5]{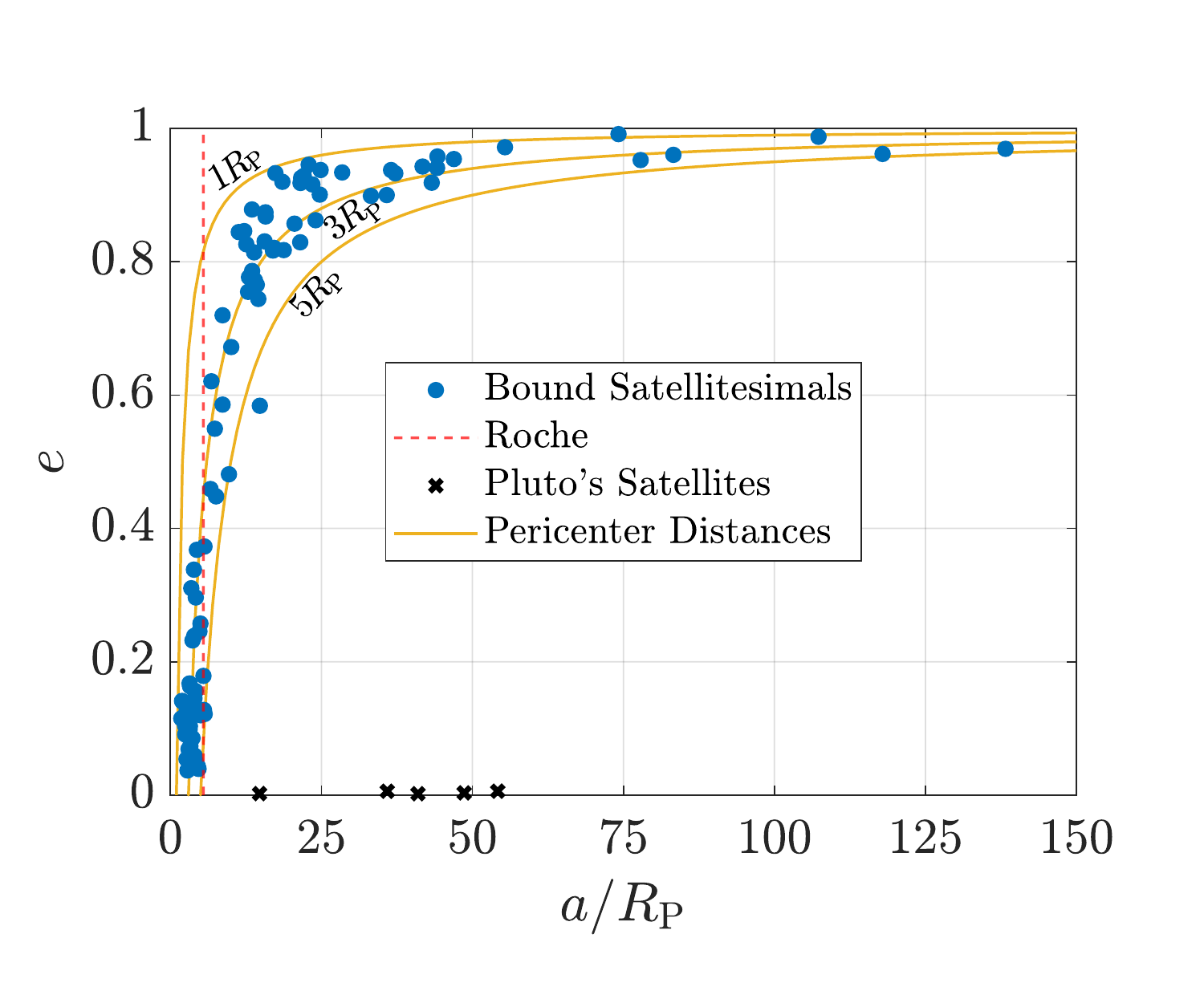}
  %\vspace{-1cm}
  \caption{Orbital elements of bound satellitesimals and of the satellites in the Plutonian system: Charon, Styx, Nix, Kerberos and Hydra \citep{kenyon2019pluto}. Contours indicate constant pericenter values.} 
  \label{f:PlutoMoons_a_e} 
\end{figure}

The relationship between eccentricity and semi-major axis of bound satellitesimals is shown in Fig.~\ref{f:PlutoMoons_a_e}. The vast majority of satellitesimals have semi-major axis lower than 25R$_{\rm P}$ and pericenter distances between 1R$_{\rm P}$ and 5R$_{\rm P}$. Eccentricity increases rapidly with distance, and beyond the Roche limit there are no objects with eccentricities below $0.4$. Already at Charon's current semi-major axis, satellitesimals reach $e\sim0.75$, and satellitesimals with semi-major axes similar to Pluto's minor satellites have $e>0.9$. Given the high eccentricities and small pericenter distances, if Pluto's minor satellites formed from intact satellitesimals, our simulations show they must have undergone substantial post-impact dynamical evolution, such as proposed by \citet{walsh2015formation}.

% \input{journal_abbrev}
% \bibliography{bibliography.bib}
% \end{document}]

%*************************************************
% ELSEVIER FORMAT

\subsection*{Acknowledgments:}
This study was supported by the Helen Kimmel Center for Planetary
Science, the Minerva Center for Life Under Extreme Planetary Conditions, the Israeli Ministry of Science (\#3–13592) and the Adolf and Mary Mil Foundation.

%% New version of the num-names style
%http://cdsads.u-strasbg.fr/abs_doc/aas_macros.html

\def\nat{Nature}
\def\ssr{Space Sci. Rev.}
\def\icarus{Icarus}
\def\jgr{J. Geophys. Res.}
\def\epsl{Earth Planet. Sci. Lett.}
\def\mps{Meteorit. Planet. Sci.}
\def\science{Science}
\def\ijie{Int. J. Impact Eng.}
\def\jgrp{J. Geophys. Res. Planets}
\def\rs{Radio Sci.}
\def\lpscp{Lunar and Planetary Science Conference Proceedings}
\def\aj{Astron. J.}
\def\sa{Sci. Adv.}
\def\aj{Astron. J.}
\def\grl{Geophys. Res. Lett.}
\def\eos{Eos (Washington DC)}
\def\rmg{Rev. Mineral. Geochem.}

%\bibliographystyle{elsarticle-num-names}
%\bibliographystyle{plain}

%%% Loading bibliography style file
\bibliographystyle{cas-model2-names}

\bibliography{bibliography.bib}

% R.I.P. Toomre
\end{document}